\documentclass[lettersize,journal]{IEEEtran}
\usepackage{mathtools,amsfonts}
\usepackage{algorithmic}
\usepackage{algorithm}
\usepackage{array}
\usepackage{textcomp}
\usepackage{stfloats}
\usepackage{url}
\usepackage{verbatim}
\usepackage{graphicx}
\usepackage{cite}
\usepackage{float}
\usepackage{amssymb}
\usepackage{caption}
\usepackage{subcaption}  
\usepackage{booktabs} 
\usepackage{tabularx} 
\allowdisplaybreaks[4] 
\begin{document}

\title{Device Association and Resource Allocation for  Hierarchical Split Federated  Learning in Space-Air-Ground Integrated Network}

\author{{Haitao Zhao, Xiaoyu Tang, Bo Xu*, Jinlong Sun, Linghao Zhang}
\thanks{The work of  was supported in part by the National Natural Science Foundation of China under Grant U2441226.}

\thanks{Haitao Zhao, Xiaoyu Tang, Bo Xu, Jinglong Sun, Linghao Zhang are with the Jiangsu Key Laboratory of Wireless Communications,  Nanjing University of Posts and Telecommunications, Nanjing 210003, China (e-mail: 
zhaoht@njupt.edu.cn, 
1025010240@njupt.edu.cn,  
xubo@njupt.edu.cn,
sunjinlong@njupt.edu.cn,
2024211102@njupt.edu.cn.
).(Corresponding author:  Bo Xu.)}

}

\maketitle

\begin{abstract}
6G facilitates deployment of Federated Learning (FL)  in the Space-Air-Ground Integrated Network (SAGIN), yet FL confronts challenges such as resource constrained and unbalanced data distribution. To address these issues, this paper proposes a Hierarchical Split Federated Learning (HSFL) framework and derives its upper bound of loss function. To minimize the weighted sum of training loss and latency, we formulate a joint optimization problem that integrates device association, model split layer selection, and resource allocation. We decompose the original problem into several subproblems, where an iterative optimization algorithm for device association and resource allocation based on brute-force split point search is proposed.  Simulation results demonstrate that the proposed algorithm can effectively balance training efficiency and model accuracy for FL in SAGIN.
\end{abstract}


\section{Introduction}

Driven by dual requirements of 6G for global coverage and edge intelligence, the Space-Air-Ground Integrated Network (SAGIN), leveraging the wide coverage of low-earth orbit (LEO) satellites, flexible edge deployment of Unmanned Aerial Vehicles (UAVs), and ubiquitous sensing of mobile devices \cite{1}, has emerged as the core architecture for scenarios such as remote communications, emergency rescue, and intelligent computing \cite{2}. In particular, Federated Learning (FL), which enables privacy-preserving collaborative global model optimization, is pivotal for SAGIN based edge intelligence \cite{4}. The authors in \cite{5}  leveraged the computation and communication resources in SAGIN, aiming to facilitate and accelerate the deployment of FL in remote regions. Besides, a hierarchical FL framework based on dynamic reward function for SAGIN was proposed in \cite{6}, using a decoupled-recoupled algorithm to optimize resource allocation and aggregation weights. However, multi-layer heterogeneity in SAGIN and resource constraints pose critical challenges to FL. On one hand, resource-constrained mobile devices cannot support efficient deep neural network (DNN) training, resulting in excessive computational latency \cite{7}. On the other hand, the heterogeneity of data collected by different mobile devices results in model aggregation error and makes it difficult for the model to converge quickly \cite{8}.

Split Federated Learning (SFL) can offer a solution to the aforementioned challenges by splitting DNN models layer-wise. Resource constrained devices handle shallow-layer computations, where deep-layer tasks are offloaded to edge servers \cite{9}. The authors in \cite{10} focused on split point selection and bandwidth allocation for minimal SFL latency, solved via alternating optimization. In  \cite{11}, a hierarchical SFL framework with convergence performance analysis was proposed,
jointly optimizing  model splitting and aggregation through iterative descent.
Besides, the authors in \cite{12} introduced a communication and computation efficient SFL framework with dynamic model splitting and aggregated gradient broadcasting.

However, existing studies have not integrated SFL with the SAGIN architecture, nor have they fully considered the impact of unbalanced data distribution, which has largely hindered the efficient training of models in the SAGIN scenario. To address these issues, we analyze the convergence performance and derive the impacts of split layer selection and device association on performance. On this basis, we establish a joint optimization problem that minimizes the weighted sum of training loss and latency by optimizing split layer selection, device association, and resource allocation. Meanwhile, in comparison with \cite{10}, we incorporate  device association modeling and, based on the analytical findings, effectively mitigate the adverse effects of imbalance data distribution on model training. Compared with \cite{13}, the proposed algorithm incorporates model spliting, which can reduce the local training load of devices. Simulation results can validate the theoretical analysis and demonstrate the effectiveness of the proposed solution.

\section{System Model}
 For the proposed SAGIN-HSFL framework, we define the sets of UAVs and mobile devices as $\mathcal{K}$ and $\mathcal{N}$, respectively. UAV $k \in \mathcal{K}$ covers $N_k$ mobile devices, and the set is denoted as $\mathcal{N}_k$.
 Besides, the set of UAVs available for association with mobile device $n\in\mathcal{N}$ is defined as $\mathcal{K}_n$. We introduce a binary variable $a_{n,k} \in \{0,1\}$ to indicate the association between device $n$ and UAV $k$, where $a_{n,k} = 1$ means device $n$ is served by UAV $k$ for local model updates, and $a_{n,k} = 0$, otherwise. The traffic load of UAV $k$ is defined as $n_k= \sum_{n \in \mathcal{N}_k} a_{n,k} $. For learning tasks, we consider classification or recognition problems involving $C$ categories, where the category set is denoted by $\mathcal{C} = \{1, 2, \dots, C\}$.

\subsection{Training Process and Convergence Analysis}
For each device $n$, the goal of model training is to find an optimal function $H_{w_n}(x)$, where $x$ is the training data, $y$ is the actual label, and $w_n$ represents the model parameters. The optimal model can be obtained by minimizing the distance between the network output and the label $y$. Therefore, the loss function of device $n$ can be formulated as
\begin{equation}
F_n\left(w_n^U, w_n^B\right) \triangleq \frac{1}{|X_n|} \sum_{x \in X_n, y \in Y_n} f\left(H_{w_n^B}\left(H_{w_n^U}(x)\right), y\right),
\end{equation}
where $w_n^U$ and $w_n^B$ denote the model parameters of the front-end and the back-end subnetwork, respectively. $|X_n|$ is the number of training data for  mobile device $n$, $x$ is the training data in set $X_n$, and $Y_n$ represents the corresponding label set of $X_n$. 

We examine  four key stages in the training process.

1) Model Initialization. At the beginning of each global training round, the satellite assesses the round's expected total training latency $T$ and its available service latency $\tau_s$. If service time is insufficient, the satellite can  select an adjacent satellite for model migration, otherwise, it initiates  FL training directly.

2) Forward Propagation. 
For the $t$-th local iteration, each device $n$ generates intermediate results $A_{n,t}^k = H_{w_{n,t}^{U,k}}(x)$ by leveraging its local data and the current shallow network parameters $w_{n,t}^{U,k}$, and subsequently uploads intermediate result and label to its associated UAV $k$ via the air-to-ground wireless link. Upon receiving the intermediate outputs from all connected mobile devices within its coverage area, UAV $k$ can perform forward propagation using the deep network parameters $w_{n,t}^{B,k}$ and compute the corresponding loss value.

3) Backward Propagation and Parameter Update. Each UAV $k$ updates the parameters of its subnetwork $w_{n,t}^{B,k}$ through backward propagation and sends the backward gradient at the split layer with
learning rate $\alpha$. Then, the model parameters are updated with gradient $\alpha \sum_{c} p^n(c) \nabla_{w_{n,t}^{U,k}} \mathbb{E}_{x|y(c)} \left\{ \log \left( H_{w_{n,t}^{U,k}}(x) \right) \right\}$ and $\alpha \sum_{c} p^n(c) \nabla_{w_{n,t}^{B,k}} \mathbb{E}_{x|y(c)} \left\{ \log \left( H_{w_{n,t}^{B,k}}(A_{n,t}^k) \right) \right\}$, respectively,
\noindent where $p^n(c)$ denotes the data distribution of device $n$, $y|(c)$ denotes the set of samples with label $y$ belonging to class $c$ in the classification task of FL.

4) Sub-model Aggregation on UAVs. After $E$ local iteration updates, all subnetworks $w_{n,t}^{B,k}$ are aggregated on the UAV, and the updated model is sent to the satellite for global aggregation. After global aggregation, the satellite distributes the generated global model to all UAVs. Then, one round of global training is completed. Edge aggregation and global aggregation are performed as
$w_{t+E+1}^k = \frac{1}{n_k} \sum_{n \in \mathcal{N}_k} a_{n,k} w_{n,t+E}^{B,k}$,
and
$w_{t+E+1}^G = \frac{1}{\sum_{k} n_k} \sum_{k} n_k w_{t+E+1}^k$, respectively.

To simplify the analysis process, we make the following assumptions.

Assumption 1: 
$F_n(w)$ is $\beta$-Lipschitz smooth with $\beta>0$, i.e., $
F_n(w) \leq F_n(w') + (w - w')^T \nabla F_n(w') + \frac{\beta}{2} \| w - w' \|_2^2.
$

Assumption 2:
$F_n(w)$ is $\mu$-strongly convex with $\mu>0$, i.e.,
$
F_n(w) \geq F_n(w') + (w - w')^T \nabla F_n(w') + \frac{\mu}{2} \| w - w' \|_2^2.
$

Assumption 3: The variance of the gradient of each layer in the device has an upper bound, i.e.,
$
\mathbb{E} \left\| \nabla F_n\left(w_{n,t}^{U,k}, w_{n,t}^{B,k}, \xi_{n,t}\right) - \nabla F_n\left(w_{n,t}^{U,k}, w_{n,t}^{B,k}\right) \right\|^2 \leq L \sigma^2,
$
where $\xi_{n,t}$ is the mini-batch data sampled from device $n$ in the $t$-th iteration, $L$ is the total number of layers of the DNN model, and $\sigma$ is a finite constant.

Assumption 4: The expected squared norm of the gradient of each layer has an upper bound, i.e.,
$
\mathbb{E} \left\| \nabla F_n\left(w_{n,t}^{U,k}, w_{n,t}^{B,k}, \xi_{n,t}\right) \right\|^2 \leq L Z^2,
$
where $Z$ is a finite constant.

Assumption 5: The gradient $\nabla_w \mathbb{E}_{x|y(c)} \left\{ \log \left( H_w(x) \right) \right\}$ is $\phi_{(c)}$-Lipschitz smooth, i.e.,
$\left\| \nabla_w \mathbb{E}_{x|y(c)} \left\{ \log \left( H_{w_1}(x) \right) \right\} - \nabla_w \mathbb{E}_{x|y(c)} \left\{ \log \left( H_{w_2}(x) \right) \right\} \right\| \leq \phi_{(c)} \left\| w_1 - w_2 \right\|$.

Assumption 6: The gradient at the $mE$-th iteration is bounded as
$
\left\| \nabla_w \mathbb{E}_{x|y(c)} \left\{ \log \left( H_{w_{n,mE}}(x) \right) \right\} \right\| \leq A_{mE},
$
where $A_{mE}$ denotes the non-negative upper bound constant of the gradient norm.

Theorem 1: Let $\rho = \frac{\beta}{\mu}$ and $\gamma = \max\{8\rho, mE\} - 1$. After $mE$ iterations, given the learning rate $\alpha = \frac{4}{\mu(\gamma + t)}$, the upper bound of the difference between the average value of the loss functions of all devices is evaluated as 
\noindent 
\begin{align}
&\mathbb{E} \left\{ \sum_{n \in \mathcal{N}_k} \frac{a_{n,k}}{n_k} \left( F_n\left( \overline{w}_{n,mE} \right)  \right) \right\} - F^*\nonumber\\\leq& \frac{\rho}{\gamma + mE} \Bigg[ \frac{8P_n}{\mu} + \frac{\mu}{2} (\gamma + 1) \mathbb{E}\sum_{n \in \mathcal{N}_k} \frac{a_{n,k}}{n_k} \left\| \overline{w}_{n,1} - w^* \right\|^2 \Bigg],
\end{align}
where $\overline{w}_{n,mE} = \left[ w_{n,mE}^{U,k}, \frac{1}{n_k} \sum_{n \in \mathcal{N}_k} a_{n,k} w_{n,mE}^{B,k} \right]$, $F^*$ is the optimal traning loss, $P_n = 2(mE\!-\!1)^2 L Z^2 \!+\! 4\beta \Gamma\! +\! \ell Z^2\! + \!\frac{1}{n_k}(L\! - \!\ell)Z^2 \!+ \!\ell \sigma^2\! + \!\frac{1}{n_k}(L\! -\! \ell)\sigma^2 \!+\! \frac{1}{2}\mu \sum_{n \in \mathcal{N}_k} \frac{1}{n_k} \mathbb{E} \{ \big( \sum_{i=0}^{k-1} A_{mE\! -\! i - \!1} \big( 1\! - \!\alpha \sum_{c} p^n(c) \phi_{(c)} \big)^i \big)^2\sum_{c} \left( a_{n,k} \| p^n(c)\! - \!p(c) \| \right)^2 \},$
and $\Gamma = \sum_{n \in \mathcal{N}_k} \frac{a_{n,k}}{n_k} (F^* - F_n^*)$ denotes the impact of data heterogeneity.

\textit{Proof.} See Proof of Theorem 1.

The theorem indicates that higher split layer $\ell$ lead to an increased convergence upper bound and degraded model performance, as fewer aggregated DNN parameters restrict the amount of device-side information, thereby weakening model generalization. Furthermore, the error upper bound escalates with the degree of data distribution discrepancy, which is strongly influenced by device association.

\subsection{Communication and Computation Models}

The path loss between UAV \( k \) and mobile device \( n \) is defined as $
PL_{n,k} = 10\varphi \log(s_{n,k}) + \eta(\omega - \omega_0)e^{\frac{\omega_0 - \omega}{\gamma}} + k_0,
$
where $ s_{n,k} $ is the horizontal distance, $ \omega $ and $ \omega_0 $ denote the vertical angle and its offset, respectively, $ \varphi $ is the terrestrial path loss exponent, $ \eta $ and $ k_0 $ represent the excess path loss and its offset, respectively, and $ \gamma $ is the angle attenuation factor \cite{14}.
Then, the uplink data rate of device \( n \) is defined as
$
r_{n,k} = a_{n,k}l_{n,k} B_k^UR_{n,k},
$
where $ B_k^U $ denotes the uplink bandwidth for the UAV $k$,  \( l_{n,k} \) is the bandwidth allocation proportion for device \( n \), \( R_{n,k}=\log_2 \left( 1 + \frac{p_n^U10^{-\frac{PL_{n,k}}{10}}}{N_0 } \right) \) is the uplink spectral efficiency, \( p_n^U \) is  the transmit power of device \( n \), and \( N_0 \) denotes the power of additive white Gaussian noise.
Thus, the transmission latency for uploading the intermediate features can be expressed as 
$
t^\ell_{n,k} = \frac{M^\ell}{r_{n,k}}
$,
where $M^\ell$ is the data size of intermediate features output by the split layer $\ell$.
In addition, when the model is split at the $\ell$-th layer, the local computation latency of device $n$ can be expressed as 
$t_{n,cp}^{\ell} = \frac{C_n^\ell}{f_n}$, where $ C_n^\ell $ represents the computational load, and $f_n$ is the computation capability of device $n$. Besides, when UAV $k$ is associated with device $n$, the computation latency at the UAV $k$ is expressed as
$
t_{k,cp}^{\ell} = \frac{C_{n,k}^\ell}{f_{n,k}}
$,
where $ C_{n,k}^\ell $ denotes the computational workload of the model part offloaded by device $ n $ to UAV $ k $, and $f_{n,k}$ is the computation resource allocated by UAV $k$ to device $n$.

Furthermore, for the UAV-satellite link, the channel gain from UAV $k$ to the satellite can be calculated by
$
H_{k,s} = G_{UAV} G_{sat} \left( \frac{\lambda}{4\pi s_{k,s}} \right)^210^{-\frac{F_{rain}}{10}}
$,
where $G_{UAV}$ and $G_{sat}$ are the antenna gains of the UAV and the satellite, respectively, $\lambda$ denotes the carrier wavelength, $s_{k,s}$ is the distance between UAV $k$ and the satellite, and $ F_{rain} $ represents the rain attenuation, which is modeled as a Weibull distribution \cite{15}.
Consequently, the transmission rate from UAV $k$ to the satellite is defined as
$
r_{k,s} = B_k^U \log_2 \left( 1 + \frac{p_k^U  H_{k,s}}{N_0} \right)
$,
where $p_k^U$ is the transmit power of the UAV $k$.
Thus, the transmission latency for uploading the edge models is determined by the bottleneck UAV and can be expressed as 
$
t_s = \max_k \frac{\bar{M^\ell}}{r_{k,s}}
$,
where $\bar{M^\ell}$ is the amount of model data uploaded by the UAV.

Finally, considering synchronous aggregation,
we further define $t_d =\max \{a_{n,k} \left( t_{n,cp}^{\ell} + t^\ell_{n,k} \right)\}$ as the maximum computation and upload latency of the device. Similarly, define $t_u = max \{a_{n,k} t_{k}^{cp}\}$ as the maximum computation latency among the UAVs. Consequently, the total system training latency is expressed as
$
T = t_d + t_u + t_s + N_{sw} \tau_s
$,
where $ N_{sw} $ represents the number of satellite handovers, and $\tau_s$ represents the switching time between satellites.

\section{Problem Formulation and Solution Approach}
\subsection{Problem Formulation}
Leveraging the results from the convergence analysis, we formulate a joint optimization problem $ P_1 $ to minimize training loss and system latency
\begin{align}
\min_{\{\ell, a_{n,k}, l_{n,k}, B_k, f_{n,k}\}} \quad &(1-\theta)T + \theta P \label{eq:P1_objective} \\
\text{s.t.}\quad\quad\quad\quad 
&\sum_{n \in \mathcal{N}_k}a_{n,k}l_{n,k}\leq1,\,k\in \mathcal{K} ,\tag{3a} \\
&\sum_{k\in\mathcal{K}} B_k^U \leq B^U, \tag{3b} \\
&\sum_{k \in \mathcal{K}_n} a_{n,k} = 1, \tag{3c} \\
&a_{n,k} \in \{0,1\}, n\in\mathcal{N}, k\in \mathcal{K} \tag{3d} \\
&\sum_{n \in \mathcal{N}_k} f_{n,k} \leq f_k, k\in \mathcal{K}\tag{3e} \\
&f_{n,k} \geq 0, n\in\mathcal{N}, k\in \mathcal{K} \tag{3f}\\
&\ell\in\{1,2,...,L\}, \tag{3g}
\end{align}
where $P=\! \ell Z^2\! + \!\frac{1}{n_k}(L\! - \!\ell)Z^2 \!+ \!\ell \sigma^2\! + \!\frac{1}{n_k}(L\! -\! \ell)\sigma^2 \!+\!\frac{1}{n_k}\sum_{c} \left(\sum_{n \in \mathcal{N}_k} a_{n,k} \| p^n(c) - p(c) \| \right)$, \( B^U \) denotes the total uplink bandwidth of all UAVs and $f_k$ represents the total computation resources of UAV $k$. 
Besides, constraints (3a) and (3b) are related to bandwidth resource allocation, constraints (3c) and (3d) pertain to device association, and constraints (3e) and (3f) are associated with computational resource allocation.

Given the split point \( \ell \), the optimization objective can be written as 
$
I=\frac{1}{n_k}\theta \sum_{c} q_{k,c} + (1-\theta)T
$,
then the subproblem $ P_2 $ is defined as
\begin{align}
\min_{\{a_{n,k}, l_{n,k}, B_k, f_{n,k}\}} \quad& \ I\\
\text{s.t. }\quad\quad\,\,\quad&\text{(3a) - (3g) }. \nonumber
\end{align}

This problem is a mixed-integer nonlinear programming (MINLP) problem. It is noteworthy that the data heterogeneity metric only depends on the association variable \( a_{n,k} \) and is decoupled from the resource allocation variables, which allows the problem to be decomposed.
\subsection{Resource Allocation}

Given the device association strategy, the problem is reformulated as minimizing latency in the device-UAV phase. Moreover, we observe that bandwidth resource allocation directly influences \( t_d \), whereas computational resource allocation independently affects \( t_u \), and these two factors are mutually independent.

\subsubsection{Bandwidth Resource Allocation}
To minimize the latency \( t_d \), we jointly optimize the bandwidth allocation ratio and UAV uplink bandwidth, i.e.,
\begin{align}
\min_{\{l_{n,k}, B_k, t_d\}} \quad &t_d \label{eq:bandwidth_subproblem} \\
\text{s.t.}\quad \,\,\, \quad &\text{(3a), (3b), (3g),}\nonumber\\
&a_{n,k} \left( \frac{C_n^\ell}{f_n} + \frac{M^\ell}{r_{n,k}} \right) \leq t_d, n\in\mathcal{N}, k\in \mathcal{K}. \tag{5a}
\end{align}
The optimal bandwidth resource allocation can be derived from the following theorem

\noindent \textbf{Theorem 2}: Given the device association \( a_{n,k} \), the optimal bandwidth allocation ratio \( l_{n,k}^* \) and UAV uplink bandwidth \( B_k^{U*} \) are expressed as
\begin{align}
l_{n,k}^* = a_{n,k}  \frac{M^\ell}{\left(t^*_d - \frac{C_n^\ell}{f_n} \right)  B_k^{U*}  R_{n,k} },
\end{align}
and
\begin{align}
B_k^{U*} = \sum_{n \in \mathcal{N}_k} a_{n,k}  \frac{M^\ell}{\left(t^*_d - \frac{C_n^\ell}{f_n} \right)   R_{n,k}},
\end{align}
where \(t^*_d \) is the optimal learning latency for local device computation and upload, which can be solved from the following equation using the bisection method i.e.,
\begin{align}
\sum_{k\in\mathcal{K}} \sum_{n\in\mathcal{N}} a_{n,k}  \frac{M^\ell}{\left(t^*_d - \frac{C_n^\ell}{f_n} \right)   R_{n,k}} = B^U.
\end{align}
\subsubsection{Computational Resource Allocation}
To minimize the UAV computational latency \( t_u \), we can derive that
\begin{align}
\min_{\{f_{n,k}\}} \quad&\ t_u\\
\text{s.t. }\quad&\text{(3e), (3f), (3g). }\nonumber
\end{align}

Since the objective function and constraints are convex, there exists a unique global optimal solution. By applying the \text{Karush-Kuhn-Tucker (KKT) conditions}, the optimal computational resource allocation is evaluated as
\begin{align}
f_{n,k}^* = \frac{C_{n,k}^\ell  f_k}{\sum_{n \in \mathcal{N}_k} C_{n,k}^\ell}.
\end{align}

\subsection{Device  Association}
After determining the optimal resource allocation, the problem is simplified to optimizing the device-UAV association to balance latency and data heterogeneity, i.e.,
\begin{align}
\min_{\{a_{n,k}\}} \quad &\theta \!\frac{1}{n_k}\sum_{c} \left(\sum_{n \in \mathcal{N}_k} a_{n,k} \| p^n(c) - p(c) \| \right)  + (1-\theta)T \label{eq:association_subproblem} \\
\text{s.t. }\quad&\text{(3c), (3d), (3g),}\nonumber\\
&\sum_{{k}\in\mathcal{K}} \sum_{{n\in\mathcal{N}}} a_{n,k}  \frac{M^\ell}{\left(t^*_d - \frac{C_n^\ell}{f_n} \right)   R_{n,k}} \leq B^U, \tag{11a} \\
&\frac{\sum_{n \in \mathcal{N}_k} C_{n,k}^\ell}{t_u^*} \leq f_k,k\in\mathcal{K}, \tag{11b}
\end{align}
where \(t^*_d \) and \( t_u^* \) are the optimal latencys for local device computation/upload and UAV computation under a given association strategy, respectively. To linearize the norm term \( \| p^n(c) - p(c) \| \), an auxiliary variable \( q_{k,c} \) is introduced, satisfying
\begin{align}
\min_{\{a_{n,k}, q_{k,c}\}} &\frac{1}{n_k}\theta \sum_{c} q_{k,c} + (1-\theta)T\\
\text{s.t. }\quad&\text{(3c), (3d), (11a), (11b), }\nonumber\\
&q_{k,c} \geq \sum_{n} a_{n,k}  (p^n(c) - p(c)),k\in\mathcal{K},c\in\mathcal{C},\tag{12a}\\
&q_{k,c} \geq \sum_{n} a_{n,k}  (p(c) - p^n(c)),k\in\mathcal{K},c\in\mathcal{C}.\tag{12b}
\end{align}

By integrating constraints into the objective function via Lagrangian relaxation , the final device association criterion is derived as:
\begin{align}
k_n^* &= \arg \min_{k \in \mathcal{K}_n} \left\{ \psi^*  \frac{M^\ell}{\left( t^{'}_d - \frac{C_n^\ell}{f_n} \right)  R_{n,k}} \right. \label{eq:association_criterion}\nonumber \\
&\quad + \nu^*  \frac{\sum_{n \in \mathcal{N}_k} C_{n,k}^\ell}{t_u'} 
 \left. + \sum_{c} \left( \lambda_{k,c}^* - \mu_{k,c}^* \right)  (p^n(c) - p(c)) \right\},
\end{align}
where  \( \lambda_{k,c}, \mu_{k,c}, \psi^*, \nu^* \) are non-negative multipliers.
This criterion comprehensively considers communication cost, computational cost, and data heterogeneity.

We adopt exhaustive search for model split layer \( l \) selection, solving the joint optimization problem over all candidate split layers and selecting the optimal layer and solution via objective function value comparison. The proposed algorithm is summarized as Algorithm 1. The algorithm traverses $L$ candidate split layers, contributing a complexity of $O(L)$. For each split layer, $T_{\text{iter}} = \max\left\{\left\lceil\frac{t_{\max}^n - t_{\min}^n}{\delta}\right\rceil, \left\lceil\frac{t_{\max}^k - t_{\min}^k}{\varepsilon}\right\rceil\right\}$ iterations are performed to optimize device association and resource allocation. In each iteration, all possible device–UAV association pairs are evaluated to derive the optimal association and corresponding resource allocation, incurring a computational cost of $O(NK)$. Moreover, bandwidth allocation by the bisection method and closed-form KKT-based resource computation incur only constant-time overhead, where the overall time complexity of the algorithm is $O(L \cdot T_{\text{iter}} \cdot NK)$.

\begin{algorithm}[!t]
\caption{Primal-Dual Algorithm for the Split Layer-Device-Resource Joint Optimization}
\label{alg:primal_dual_joint}
\textbf{Input}: $[t_d^{\text{min}}, t_d^{\text{max}}], [t_u^{\text{min}}, t_u^{\text{max}}]$, total bandwidth $B^U$, computation capability $f_n$ and $f_k$, total number of layers $L$.

\textbf{Output}: Optimal split layer $\ell^*$, device association $a_{n,k}^*$, resource allocation $l_{n,k}^*, B_k^{U*}, f_{n,k}^*$.
\begin{enumerate}
\renewcommand{\labelenumi}{\arabic{enumi}:}
\item Initialize device association $a_{n,k}$, resource allocation $l_{n,k}, B_k, f_{n,k}$, objective $I$, $P = +\infty$.
\item \textbf{for} $\ell = 1$ to $L$:
\item \quad  Initialize $t_d = t_d^{\text{max}}$, $t_u = t_u^{\text{max}}$, $I^* = +\infty$.
\item \quad \textbf{repeat}
\item \quad \quad Solve the problem under the conditions of $t_d$ and $t_u$ to obtain the optimal device association matrix $a_{n,k}$.
\item \quad \quad Obtain $I$, $t_d$ and $t_u$ through $a_{n,k}$. 
 \item\quad \quad Set $t_d = t_d - \delta$ and $t_u = t_u - \varepsilon$ (where $\delta,\varepsilon>0$).
\item \quad \quad \textbf{if} $I < I^*$, set $I^* = I$, $a_{n,k}^{*} = a_{n,k}$.\textbf{end if}
\item \quad \textbf{until} $t_d \leq t_d^{\text{min}}$ and $t_u \leq t_u^{{\text{min}}}$.
\item \quad  Obtain the optimal device association $a_{n,k}^*$.
\item \quad According to Theorem 2 and $a_{n,k}^*$, obtain the resource allocation scheme $l_{n,k}^*$, $B_k^{U*}$ , $f_{n,k}^*$ , $P^*$.
\item \quad  \textbf{if} $P^{*} < P$, update $P \leftarrow P^{*}$, $\ell \leftarrow \ell^{*}$, $a_{n,k} \leftarrow a_{n,k}^{*}$, $l_{n,k} \leftarrow l_{n,k}^{*}$, $B_k^{U} \leftarrow B_k^{U*}$, $f_{n,k} \leftarrow f_{n,k}^{*}$.\textbf{end if}
\item \textbf{end for}
\end{enumerate}
\end{algorithm}

\section{Simulation and Analysis}
In the simulation experiments, we consider a SAGIN scenario consisting of 4 UAVs and 50 mobile devices, where the devices are randomly distributed within the target area. The coverage range and altitude of the UAVs are 1500m and 500m. The transmit power of the mobile devices is 28 dBm, the noise power spectral density is -174 dBm/Hz, and the total uplink bandwidth is 50 MHz. Regarding model training, the local iteration count is 5, the batch size is 64, the global iteration count is 2000, and the learning rate is 0.02. Each device has one or more UAVs available for association. AlexNet is trained on the CIFAR-10 dataset. Training data is evenly distributed across all devices, with the local dataset of each device containing only 4 of the 10 classes to ensure non-independent and identically distributed (non-IID) data across devices.

In this paper, MATLAB’s \texttt{walkerStar} and \texttt{accessIntervals} functions are used to construct the constellation model and calculate each satellite’s coverage time over the target area, respectively \cite{5}. Eighty LEO satellites are evenly distributed across 10 orbits, with an orbital altitude of 800 km and an inclination of 85°. The minimum communication elevation angle is set to 15°, and the target area is located at 40° N latitude and 86° W longitude. The satellite closest to the target area is selected for training in each round.

To validate the effectiveness of the proposed algorithm, we employ two key evaluation metrics, namely, test accuracy and training latency, to conduct a comprehensive comparative analysis against four representative baseline algorithms: \textbf{(1) Random Association (RA) [9]}: Based on the SFL framework, it employs random device association and optimizes resource allocation with model splitting. \textbf{(2) Equal Resource Allocation (ERA)}: Performs device association based on maximum SNR and allocates resources equally. \textbf{(3) Hierarchical Federated Learning (HFL) [12]}: Adopts the HFL for association and resource optimization, explicitly excluding model splitting. \textbf{(4) Data Distribution-based Association (DDA)}: Determines device association based on data distribution metrics, with simultaneous resource allocation.
\begin{figure}[H]
    \vspace{-10pt} 
    \centering  
    \begin{subfigure}[b]{0.49\linewidth}
        \centering
        \includegraphics[width=\linewidth]{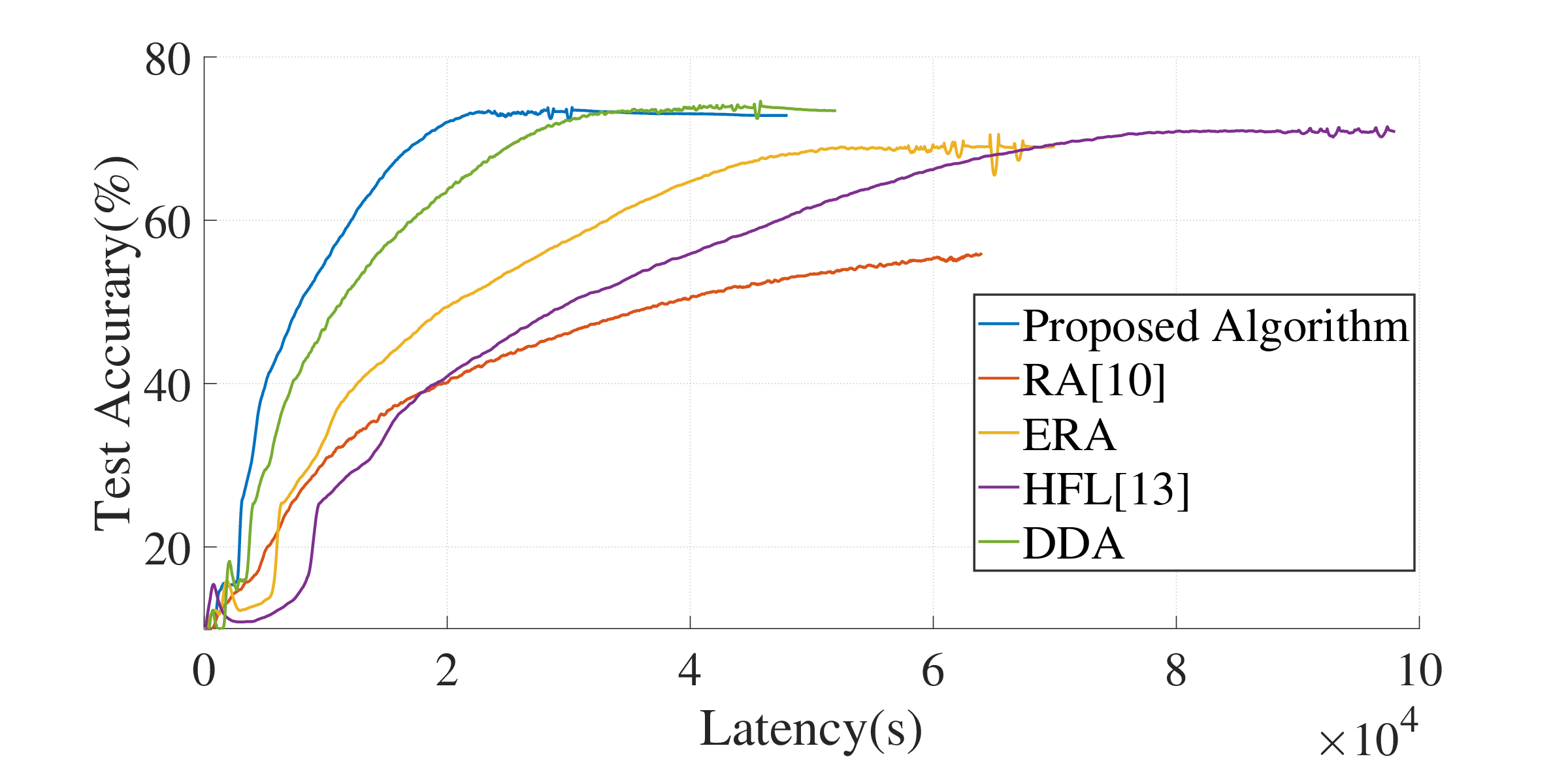}  
        \subcaption{Test Accuracy}  
        \label{fig:alexnet_test_acc}  
    \end{subfigure}
    \hfill  
    \begin{subfigure}[b]{0.49\linewidth}
        \centering
        \includegraphics[width=\linewidth]{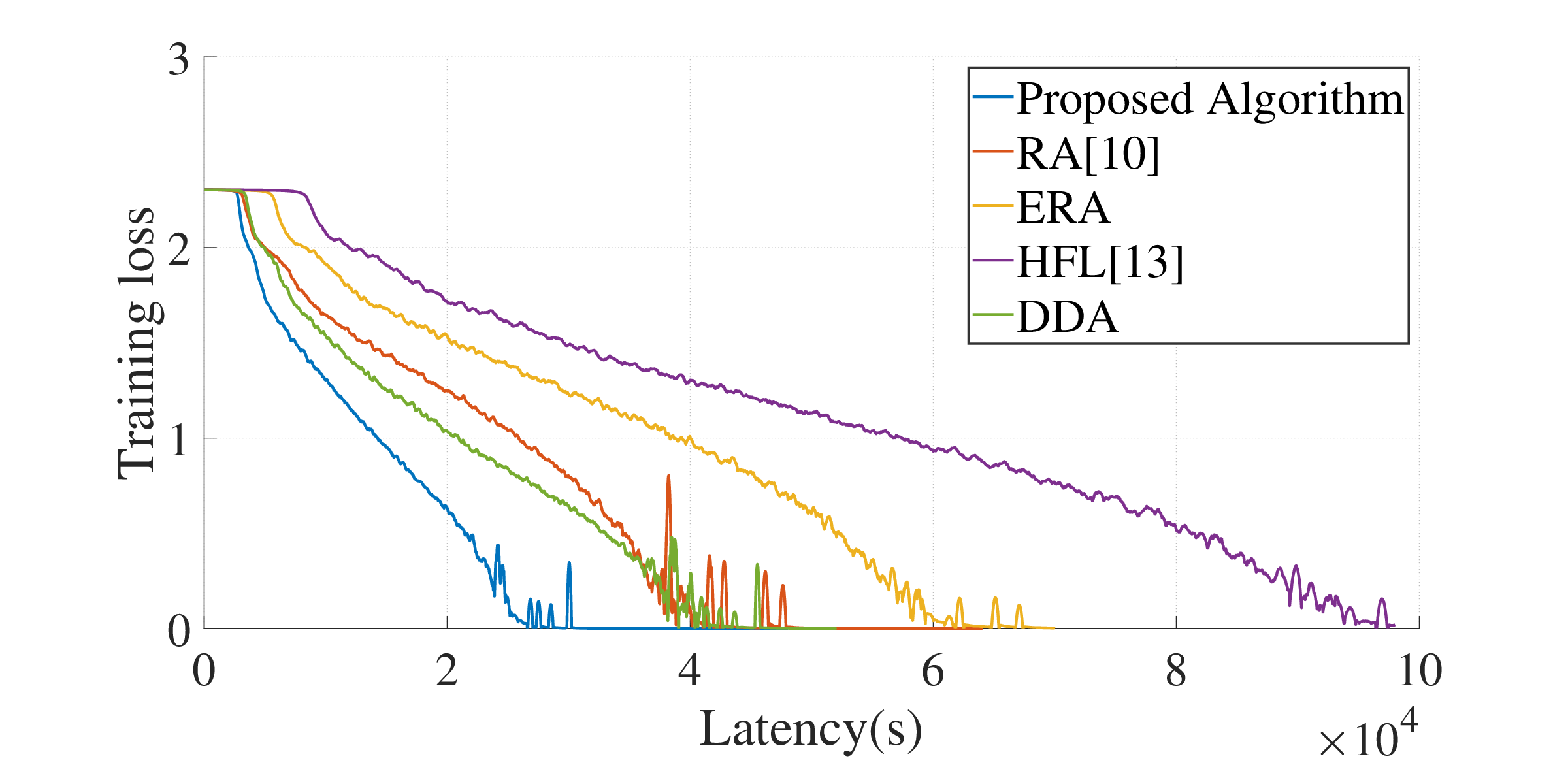}  
        \subcaption{Training Loss}  
        \label{fig:alexnet_train_loss}  
    \end{subfigure}
    \caption{Learning performance of AlexNet on CIFAR-10}  
    \label{fig:alexnet_performance}  
    \vspace{-15pt} 
\end{figure}
Fig. \ref{fig:alexnet_performance} contrasts learning performance across different device association and resource allocation strategies. Specifically, Fig. 1(a) and 1(b) show AlexNet’s test accuracy and training loss on the CIFAR-10 dataset, respectively. Fig. 1(a) demonstrates that our proposed algorithm reduces learning latency and enhances training accuracy over the same period compared to baselines, while Fig. 1(b) shows it also achieves lower training loss. Compared with the optimal baseline (DDA), the proposed algorithm improves the final test accuracy by approximately $2.7\%$ and reduces the convergence latency by around $23\%$, exhibiting superior overall performance. This is attributed to SFL mitigating resource constraint impacts, optimized device association reducing latency and alleviating non-IID data effects, and resource allocation cutting latency across all learning stages.

\begin{center}
\vspace{-15pt}
    \begin{minipage}{0.24\textwidth} 
        \begin{figure}[H]
            \centering
            \includegraphics[width=\linewidth]{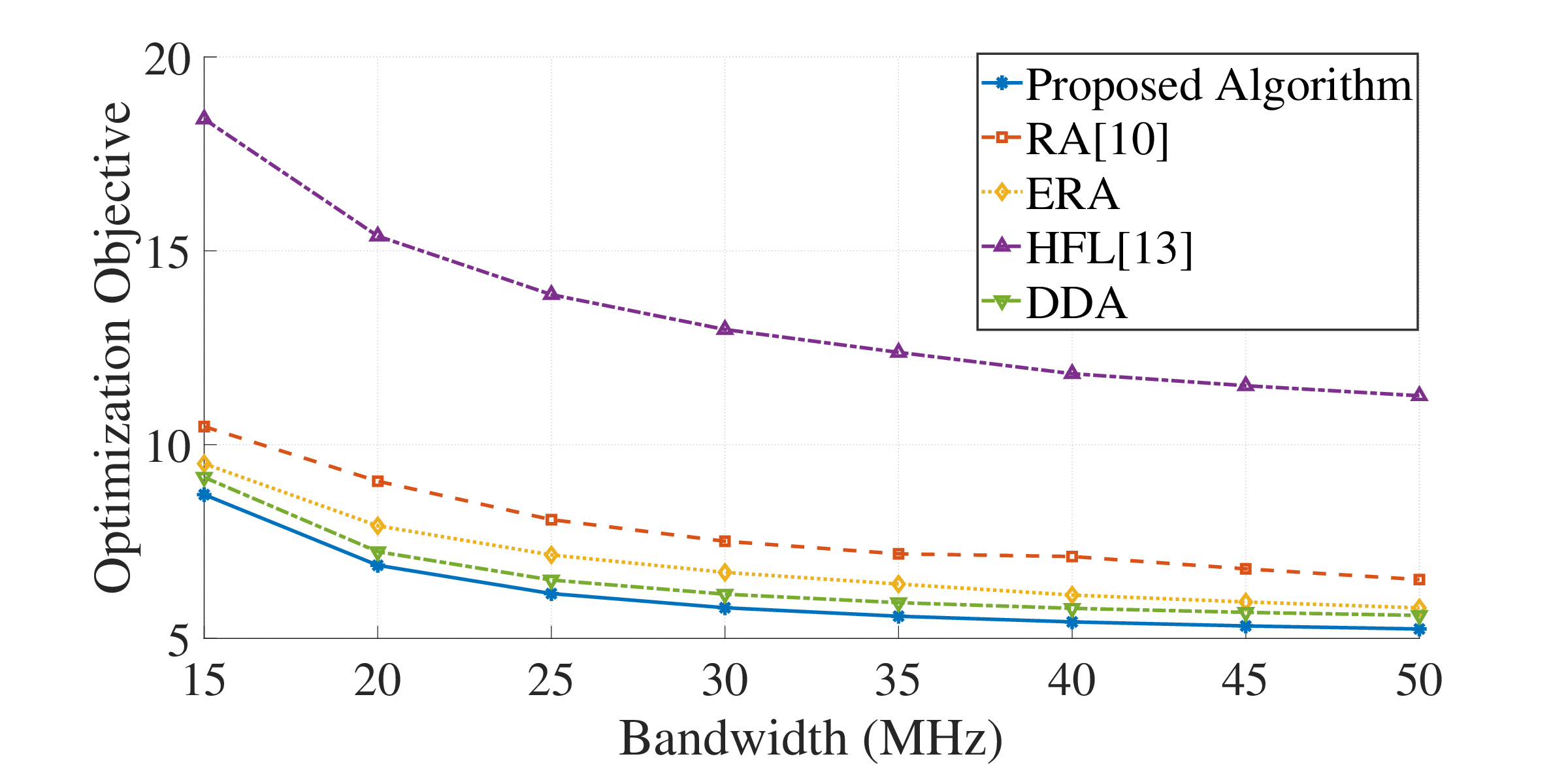}
            \caption{The impact of bandwidth}
            \label{fig:bandwidth_objective}
        \end{figure}
    \end{minipage}
    \hfill 
    \begin{minipage}{0.24\textwidth} 
        \begin{figure}[H]
            \centering
            \includegraphics[width=\linewidth]{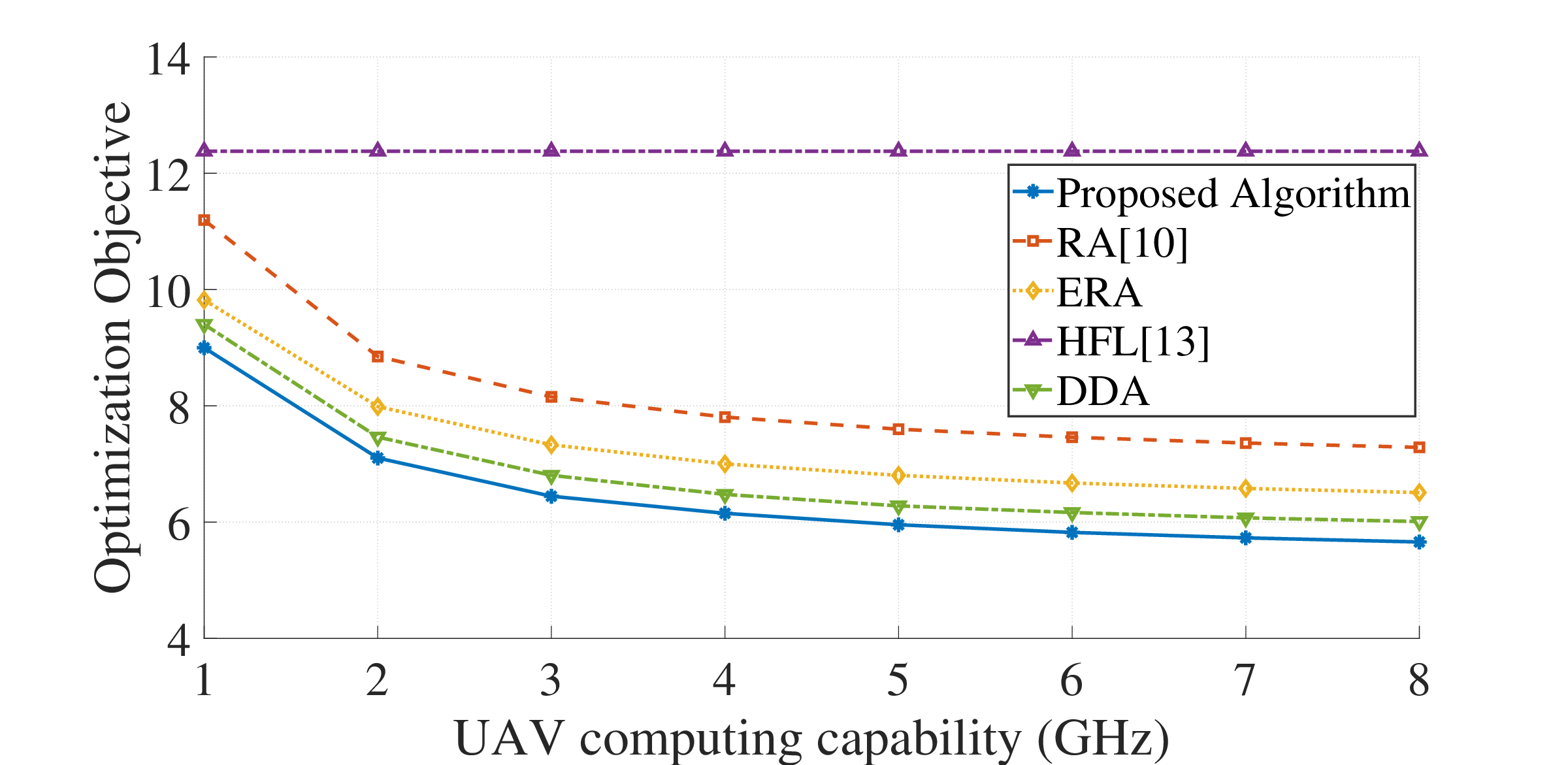}
            \caption{The impact of UAV computing capability}
            \label{fig:uav_computing_latency}
        \end{figure}
    \end{minipage}
\end{center}

Fig. \ref{fig:bandwidth_objective} compares optimization objective values across varying total uplink bandwidths. The objective value decreases as total uplink bandwidth increases, as additional bandwidth allocated to devices reduces model upload communication latency, in turn lowering the objective value.

Fig. \ref{fig:uav_computing_latency} compares training latencies across varying UAV computing capabilities. Clearly, the optimization objective value decreases as UAV computing capability improves, which is because more computing resources allocated to associated devices reduce UAV-side computing latency, in turn lowering the objective value. Notably, the HFL Algorithm without UAV-assisted computing maintains a constant objective value.

\section{Conclusion}
In this paper, we addressed device heterogeneity, unbalanced data distribution, and high latency of FL in SAGIN by proposing a HSFL framework. We formulated a joint optimization problem integrating split layer selection, device association, and resource allocation, and developed an alternating iterative algorithm to solve this MINLP problem. Simulations on the CIFAR-10 dataset with AlexNet demonstrated that the method outperformed four baselines in test accuracy and training latency. In future work, we will explore dynamic split layer adjustment to adapt to time-varying network conditions.

\section*{Proof of Theorem 1}
For the convenience of proof, we annotate some variables in the training process. We define two additional variables $v_{n,t+1}^{U,k}$ and $v_{n,t+1}^{B,k}$ to be the intermediate result of one iteration update of $w_{n,t}^{U,k}$ and $w_{n,t}^{B,k}$. Therefore, we have the following results:
\begin{small}
\begin{equation}
\begin{cases}
v_{n,t+1}^{U,k} = w_{n,t}^{U,k} - \alpha \sum_{c} p^n(c) \nabla_{w_{n,t}^{U,k}} \mathbb{E}_{x|y(c)} \left\{ \log \left( H_{w_{n,t}^{U,k}}(x) \right) \right\}; \\
v_{n,t+1}^{B,k} = w_{n,t}^{B,k} - \alpha \sum_{c} p^n(c) \nabla_{w_{n,t}^{B,k}} \mathbb{E}_{x|y(c)} \left\{ \log \left( H_{w_{n,t}^{B,k}}(A_{n,t}^k) \right) \right\}.
\end{cases}
\end{equation}
\end{small}
and
\begin{small}
\begin{equation}
w_{n,t+1}^{U,k} = v_{n,t+1}^{U,k}, \nonumber
\end{equation}
\begin{equation}
w_{n,t+1}^{B,k}=
\begin{cases}
v_{n,t+1}^{B,k} & \text{if } t+1 \notin mE ;\\
\sum_{n\in\mathcal{N}_k}\frac{a_{n,k}}{n_k}v_{n,t+1}^{B,k} & \text{if } t+1 \in mE.
\end{cases}
\end{equation}
\end{small}

To further simplify the analysis, we define the total parameter of device \( k \) as \( w_n = [w_n^U, w_n^B] \), and we have \( \overline{v}_{n,t} = [v_{n,t}^{U,k}, \sum_{n\in\mathcal{N}_k}\frac{a_{n,k}}{n_k}v_{n,t}^{B,k}] \), \( \overline{w}_{n,t} = [w_{n,t}^{U,k}, \sum_{n \in \mathcal{N}_k}\frac{a_{n,k}}{n_k}  w_{n,t}^{B,k} ] \).  
Additionally, we define the component-wise gradients for clarity:
The gradient of model parameters deployed on device is denoted as:
\begin{equation}
  \nabla F_n(w_{n,t})_U = \sum_{c} p^n(c) \nabla_{w_{n,t}^{U,k}} \mathbb{E}_{x|y(c)} \left\{ \log \left( H_{w_{n,t}^{U,k}}(x) \right) \right\}.
\end{equation}
The gradient of model parameters deployed on UAV is denoted as:
\begin{equation}
  \nabla F_n(w_{n,t})_B = \sum_{c} p^n(c) \nabla_{w_{n,t}^{B,k}} \mathbb{E}_{x|y(c)} \left\{ \log \left( H_{w_{n,t}^{B,k}}(A_{n,t}^k) \right) \right\}.
\end{equation}
Then, the gradient of model parameter is denoted as:
\begin{equation}
g_{n,t} = \left[ \nabla F_n(w_{n,t}, \xi_{n,t})_U, \sum_{n \in \mathcal{N}_k}\frac{a_{n,k}}{n_k} \nabla F_n(w_{n,t}, \xi_{n,t})_B \right]\nonumber,
\end{equation}
\begin{equation}
\overline{g}_{n,t} = \left[ \nabla F_n(w_{n,t})_U, \sum_{n \in \mathcal{N}_k}\frac{a_{n,k}}{n_k} \nabla F_n(w_{n,t})_B \right].
\end{equation}
Therefore, \( \overline{v}_{n,t+1} = \overline{w}_{n,t} - \alpha g_{n,t} \), and \( \mathbb{E}g_{n,t} = \overline{g}_{n,t} \).

We analyze the expected average difference between the device's parameter and the optimal parameter $\mathbb{E}\left(\sum_{n \in \mathcal{N}_k}\frac{a_{n,k}}{n_k}\left\|\overline{v}_{n,t+1} - w^*\right\|^2\right)$ as follows:
\begin{align}
& \mathbb{E} \sum_{n \in \mathcal{N}_k}\frac{a_{n,k}}{n_k}\left\| \overline{v}_{n, t+1}-w^{*}\right\| ^{2} \nonumber\\
&= \mathbb{E} \sum_{n \in \mathcal{N}_k}\frac{a_{n,k}}{n_k}\left\| \overline{w}_{n, t}-\alpha g_{n, t}-w^{*}-\alpha \overline{g}_{n, t}+\alpha \overline{g}_{n, t}\right\| ^{2} \nonumber\\
&= \mathbb{E} \sum_{n \in \mathcal{N}_k}\frac{a_{n,k}}{n_k}\left(\underbrace{\left\| \overline{w}_{n, t}-w^{*}-\alpha\overline{g}_{n, t}\right\| ^{2}}_{A_{1}} \right.\nonumber\\
&\quad+\underbrace{2 \alpha\left\langle\overline{w}_{n, t}-w^{*}-\alpha \overline{g}_{n, t}, \overline{g}_{n, t}-g_{n, t}\right\rangle}_{A_{2}} \nonumber\\
&\quad\left.+\underbrace{\alpha^{2}\left\| g_{n, t}-\overline{g}_{n, t}\right\| ^{2}}_{A_{3}}\right) .
\end{align}

Notice that we have $\mathbb{E}g_{n,t} = \overline{g}_{n,t}$ in the previous analysis, therefore $\mathbb{E}A_2 = 0$. We know that $A_1$ and $A_3$ satisfies the following property \cite{10}:
\begin{align} 
A_{1}= &  \sum_{n \in \mathcal{N}_k}\frac{a_{n,k}}{n_k}\left\| \overline{w}_{n, t}-w^{*}-\alpha \overline{g}_{n, t}\right\| ^{2} \nonumber\\ 
\leq & \left(1-\frac{1}{2}\mu \alpha\right)  \sum_{n \in \mathcal{N}_k}\frac{a_{n,k}}{n_k}\left\| \overline{w}_{n, t}-w^{*}\right\| ^{2} \nonumber\\ 
&-\frac{1}{2}\mu \alpha \sum_{n \in \mathcal{N}_k}\frac{a_{n,k}}{n_k}\left\| {w}_{n, t}-w^{*}\right\| ^{2}\nonumber\\
& +2  \sum_{n \in \mathcal{N}_k}\frac{a_{n,k}}{n_k}\left\| \overline{w}_{n, t}-w_{n,t}\right\| ^{2}\nonumber\\
&+\alpha^2(\ell Z^2+\frac{1}{n_k}(L-\ell)Z^2)+4 \alpha^{2}\beta  \Gamma . 
\end{align}    
\begin{align} 
\mathbb{E} A_{3}= & \mathbb{E} \sum_{n \in \mathcal{N}_k}\frac{a_{n,k}}{n_k}\left\| g_{n,t}-\overline{g}_{n,t}\right\| ^{2} \nonumber\\ 
= & \sum_{n \in \mathcal{N}_k}\frac{a_{n,k}}{n_k}\left\| \left[\nabla F_{k}\left(w_{n,t}, \xi_{n,t}\right)\right]_{U}-\left[\nabla F_{k}\left(w_{n,t}\right)\right]_{U}\right\| ^{2}\nonumber \\ 
& +\sum_{n \in \mathcal{N}_k}\frac{a_{n,k}}{n_k}\left\| \sum_{n \in \mathcal{N}_k}\frac{a_{n,k}}{n_k}\left(\left[\nabla F_{k}\left(w_{n,t}, \xi_{n,t}\right)\right]_{B}\right. \right.\nonumber\\
& \left. \left.-\left[\nabla F_{k}\left(w_{n,t}\right)\right]_{B}\right)\right\| ^{2} \nonumber\\ 
= & \sum_{n \in \mathcal{N}_k}\frac{a_{n,k}}{n_k}\left\| \left[\nabla F_{k}\left(w_{n,t}, \xi_{n,t}\right)\right]_{U}-\left[\nabla F_{k}\left(w_{n,t}, \xi_{n,t}\right)\right]_{U}\right\| ^{2} \nonumber\\ 
& +\sum_{n \in \mathcal{N}_k}\frac{a_{n,k}}{n_k}\left\| \left[\nabla F_{k}\left(w_{n,t}, \xi_{n,t}\right)\right]_{B}-\left[\nabla F_{k}\left(w_{n,t}\right)\right]_{B}\right\| ^{2} \nonumber\\ 
\leq & \ell \sigma^{2}+\frac{1}{n_k}(L-\ell) \sigma^{2} . 
\end{align}

To bound $\sum_{n \in \mathcal{N}_k}\frac{a_{n,k}}{n_k}\|\overline{w}_{k,t}^B - w_{k,t}^B\|^2$, we first assume $E$ steps between two aggregations. Therefore, for any $t \geq 0$, there exists a $t_0 \leq t$, such that $t - t_0 \leq E - 1$ and $\overline{w}_{n,t_0}^B = w_{n,t_0}^B$, then,
\begin{align}
&\mathbb{E}\sum_{n \in \mathcal{N}_k}\frac{a_{n,k}}{n_k}\|\overline{w}_{n,t} - w_{n,t}\|^2\nonumber\\
&= \mathbb{E}\sum_{n \in \mathcal{N}_k}\frac{a_{n,k}}{n_k}\left\|({w}_{n,t} - \overline{w}_{n,t_0}) - (\overline{w}_{n,t} - \overline{w}_{n,t_0})\right\|^2 \nonumber\\
&\leq \mathbb{E}\sum_{n \in \mathcal{N}_k}\frac{a_{n,k}}{n_k}\|w_{n,t} - \overline{w}_{n,t_0}\|^2 \nonumber\\
&\leq \sum_{n \in \mathcal{N}_k}\frac{a_{n,k}}{n_k}\mathbb{E}\sum_{t=t_0}^t (E-1)\alpha^2\|\nabla F_n(w_{n,t},\xi_{n,t})\|^2 \nonumber\\
&\leq (E-1)^2\alpha^2LZ^2,
\end{align}
where the first inequality is due to $\mathbb{E}\|X - \mathbb{E}X\|^2 \leq \mathbb{E}\|X\|^2$, and the second inequality is due to Jensen inequality:
\begin{align}
\|w_{n,t} - \overline{w}_{n,t_0}\|^2
&= \left\|\sum_{t=t_0}^t \alpha \nabla F_k(w_{n,t},\xi_{n,t})\right\|^2 \nonumber\\
&\leq (t - t_0)\sum_{t=t_0}^t \alpha^2\|\nabla F_k(w_{n,t},\xi_{n,t})\|^2.
\end{align}

Further more, we analysis $\| {w}_{n, t}-w^{*}\|$. When $t\in mE$, with assumption 5 and 6 \cite{13}, we have:
\begin{align}
\| {w}_{n, mE}-w^{*}\|&\leq \left\|{w}_{n,mE-1} - {w}^{*} \right\| \nonumber\\
&\quad - \alpha \left\| \sum_{c} p^k(c) \nabla_{{w}} \mathbb{E}_{x|y(c)} \left\{ \log \left( \mathcal{H}_{{w}_{n,mE-1}} (x) \right) \right\} \right. \nonumber\\
&\quad \left. - \sum_{c} p(c) \nabla_{{w}} \mathbb{E}_{x|y(c)} \left\{ \log \left( \mathcal{H}_{{w}^{*}} (x) \right) \right\} \right\|\nonumber\\
&\leq \left( 1 - \alpha \sum_{c} p(c) L(c) \right) \left\| {w}_{n,mE-1} - {w}^{*} \right\|\nonumber\\
&\quad - \alpha A_{mE-1} \sum \left\| p^k(c) - p(c) \right\|.
\end{align}
where $p^k(c)$ is the data distribution of UAV $k$. Continuing the iteration until the $(m-1)E\text{-th}$ iteration, we can obtain:
\begin{small}
\begin{align}
&\left\| {w}_{n,mE} - {w}^* \right\|\nonumber\\
&\leq \left( 1 - \alpha \sum_{c} p(c) L(c) \right)^E \left\| {w}_{n,(m-1)E} - {w}^* \right\| \nonumber\\
& - \alpha \left( \sum_{i=0}^{E-1} A_{mE-1-i} \left( 1 - \alpha \sum_{c} p(c) L(c) \right)^i \right) \sum_{c} \left\|  \left( p^n(c) - p(c) \right) \right\|.
\end{align}
\end{small}

Finally, we have (26).
\begin{figure*}[t] 
\begin{equation}
\begin{aligned}
\mathbb{E} \sum_{n \in \mathcal{N}_k}\frac{a_{n,k}}{n_k}\left\| \overline{{v}}_{n,mE} - {w}^* \right\|^2
&\leq (1 - \frac{1}{2}\mu \alpha) \sum_{n \in \mathcal{N}_k}\frac{a_{n,k}}{n_k}\mathbb{E} \left\| \overline{{w}}_{n,mE} - {w}^* \right\|^2-\frac{1}{2}\mu \alpha \sum_{n \in \mathcal{N}_k}\frac{a_{n,k}}{n_k}\left\| {w}_{n, mE}-w^{*}\right\| ^{2}+ 2(mE-1)^2 \alpha^2 L Z^2 \\&\quad
+ 4\alpha^2\beta \Gamma + \alpha^2 \left( \ell Z^2 + \frac{1}{n_k}(L - \ell)Z^2 \right) + \alpha^2 \left( \ell \sigma^2 + \frac{1}{n_k}(L - \ell)\sigma^2 \right)\\
&\leq (1 - \frac{1}{2}\mu \alpha) \sum_{n \in \mathcal{N}_k}\frac{a_{n,k}}{n_k}\mathbb{E} \left\| \overline{{w}}_{n,mE} - {w}^* \right\|^2 \\
&\quad+ \frac{1}{2}\mu\alpha^2 \sum_{n \in \mathcal{N}_k} \frac{a_{n,k}}{n_k} \mathbb{E}\Bigg\{ \Bigg( \sum_{i=0}^{E-1} A_{mE-1-i} \left( 1 - \alpha \sum_{c} p(c) L(c) \right)^i \Bigg) 
  \sum_{c} \left\|\left( p^n(c) - p(c) \right) \right\| \Bigg\}^2\\
&\quad+ 2(mE-1)^2 \alpha^2 L Z^2 + 4\alpha^2\beta \Gamma + \alpha^2 \left( \ell Z^2 + \frac{1}{n_k}(L - \ell)Z^2 \right) + \alpha^2 \left( \ell \sigma^2 + \frac{1}{n_k}(L - \ell)\sigma^2 \right)\\
\end{aligned}
\end{equation}
\end{figure*}

Let $P_n = 2(mE\!-\!1)^2 L Z^2 \!+\! 4\beta \Gamma\! +\! \ell Z^2\! + \!\frac{1}{n_k}(L\! - \!\ell)Z^2 \!+ \!\ell \sigma^2\! + \!\frac{1}{n_k}(L\! -\! \ell)\sigma^2 \!+\! \frac{1}{2}\mu \sum_{n \in \mathcal{N}_k} \frac{1}{n_k} \mathbb{E} \{ \big( \sum_{i=0}^{k-1} A_{mE\! -\! i - \!1} \big( 1\! - \!\alpha \sum_{c} p^n(c) L(c) \big)^i \big)^2\sum_{c} \left( a_{n,k} \| p^n(c)\! - \!p(c) \| \right)^2 \}$, and $\Delta_t = \mathbb{E}\sum_{n \in \mathcal{N}_k} \frac{a_{n,k}}{n_k}\left\|\overline{{w}}_{n,t} - {w}^*\right\|^2$. We assume $\alpha = \frac{2\lambda}{\gamma + t}$ for some $\lambda > \frac{1}{\mu}$ and $\gamma > 0$. Now we prove $\Delta_t \leq \frac{v}{\gamma + t}$ by induction, where $v = \max\left\{ \frac{4\lambda^2 P_n}{\lambda\mu - 1}, (\gamma + 1)\Delta_1 \right\}$. Obviously, the definition of $v$ ensures it holds for $t = 1$. For some $t\in mE,$
\begin{align}
\Delta_{t+1} &\leq (1 - \mu\alpha)\Delta_t + \alpha^2 P_n \nonumber\\
&\leq \left(1 - \frac{\lambda\mu}{t+\gamma}\right) \frac{v}{t+\gamma} + \frac{4\lambda^2 P_n}{(t+\gamma)^2} \nonumber\\
&= \frac{t+\gamma - 1}{(t+\gamma)^2}v + \left[ \frac{4\lambda^2 P_n}{(t+\gamma)^2} - \frac{\lambda\mu - 1}{(t+\gamma)^2}v \right]\nonumber \\
&\leq \frac{v}{\gamma + t + 1}
\end{align}

Then by the $\beta$-smoothness of $F(\cdot),$
\[
\mathbb{E}\sum_{n \in \mathcal{N}_k} \frac{a_{n,k}}{n_k} F_n(\overline{\boldsymbol{w}}_{n,t}) - F^* \leq \frac{\beta}{2}\Delta_t \leq \frac{\beta}{2} \frac{v}{\gamma + t}
\]
Specifically, let $\lambda = \frac{2}{\mu}, \rho = \frac{\beta}{\mu}$ and $\gamma = \max\{8\rho, E\} - 1,$ then
\begin{align}
v &= \max\left\{ \frac{4\lambda^2 P_n}{\lambda\mu - 1}, (\gamma + 1)\Delta_1 \right\} \nonumber\\
&\leq \frac{4\lambda^2 P_n}{\lambda\mu - 1} + (\gamma + 1)\Delta_1 \nonumber\\
&\leq \frac{16P_n}{\mu^2} + (\gamma + 1)\Delta_1
\end{align}
Therefore, we have
\begin{equation}
\begin{aligned}
\mathbb{E}\sum_{n \in \mathcal{N}_k} \frac{a_{n,k}}{n_k} F_n(\overline{\boldsymbol{w}}_{n,t}) - F^* &\leq \frac{\beta}{2} \frac{v}{\gamma + t} \\
&\leq \frac{\rho}{\gamma + t} \left( \frac{8P_n}{\mu} + \frac{\mu(\gamma + 1)}{2}\Delta_1 \right).
\end{aligned}
\end{equation}

\bibliography{ref}
\bibliographystyle{IEEEtran}

\end{document}